\documentclass[twocolumn,prc,showpacs,preprintnumbers,superscriptaddress]{revtex4-2}

\usepackage[T1]{fontenc}
\usepackage{amsmath} 
\usepackage{amssymb}
\usepackage{amsfonts}
\usepackage{mathrsfs}
\usepackage{units}
\usepackage{multirow}
\usepackage{color}
\usepackage{braket}
\usepackage{ulem}
\usepackage{dcolumn}
\usepackage{bm}
\usepackage{rotating} 
\usepackage{orcidlink}
\usepackage{rotating,epsfig,dcolumn}
\usepackage{graphicx,bm}
\usepackage{bm,color}
\usepackage{graphicx}
\usepackage{epstopdf}
\usepackage{rotating}
\usepackage{float}
\usepackage{multirow}
\usepackage{lipsum}
\usepackage[utf8]{inputenc}

\newcommand{\disregard}[1]{}

\newcommand{\bL}{\begin{Large}}
\newcommand{\eL}{\end{Large}}
\newcommand{\be}{\begin{equation}}  
\newcommand{\ee}{\end{equation}}
\newcommand{\ba}{\begin{eqnarray*}}
\newcommand{\ea}{\end{eqnarray*}}

\newcommand{\gras}[1]{\boldsymbol{#1}}

\newcommand{\bn}{\begin{eqnarray}}
\newcommand{\en}{\end{eqnarray}}
\newcommand{\bc}{\begin{center}}
\newcommand{\ec}{\end{center}}
\newcommand{\bi}{\begin{itemize}}
\newcommand{\ei}{\end{itemize}}

\usepackage{tikz}
\newcommand{\tikzcircle}[2][red,fill=red]{\tikz[baseline=-0.5ex]\draw[#1,radius=#2] (0,0) circle ;}%

\newsavebox{\tmpstrikebox}
\newlength{\tmpstrikelen}

\begin{document}

\title{Influence of configuration-interaction on isospin impurities and isospin symmetry
breaking corrections to superallowed $0^+\rightarrow 0^+$ beta decays.}

\author{Jakub Wysocki\orcidlink{0009-0008-3507-6650}}
\email{jj.wysocki@uw.edu.pl}
\affiliation{Institute of Theoretical Physics, Faculty of Physics, University of Warsaw, ul. Pasteura 5, PL-02-093, Warsaw, Poland}

\author{Jan Miśkiewicz\orcidlink{0000-0003-1206-6157}}
\email{j.miskiewicz@uw.edu.pl}
\affiliation{Institute of Theoretical Physics, Faculty of Physics, University of Warsaw, ul. Pasteura 5, PL-02-093, Warsaw, Poland}

\author{Jagjit Singh\orcidlink{0000-0002-3198-4829}}
\email{jagjit.singh@manchester.ac.uk}
\affiliation{Department of Physics and Astronomy, The University of Manchester, Manchester M13 9PL, UK}
\affiliation{Institute of Theoretical Physics, Faculty of Physics, University of Warsaw, ul. Pasteura 5, PL-02-093, Warsaw, Poland}
\affiliation{Department of Physics, Akal University, Talwandi Sabo, Bathinda, Punjab 151302, India}
\affiliation{Research Center for Nuclear Physics (RCNP), Osaka University, Ibaraki 567-0047, Japan}

\author{Wojciech Satu{\l}a\orcidlink{0000-0003-0203-3773}}
\email{Wojciech.Satula@fuw.edu.pl}
\affiliation{Institute of Theoretical Physics, Faculty of Physics, University of Warsaw, ul. Pasteura 5, PL-02-093, Warsaw, Poland}
\

\date{\today}

\begin{abstract}{
The symmetry-conserving density functional theory (DFT)-based no-core configuration-interaction (DFT-NCCI) framework is 
applied for the first time  to investigate the impact of configuration interaction (CI) on the Coulomb (isospin) impurity, $\alpha_{\rm C}$, 
in the ground and excited states of $^{10}$C, $^{10}$B, and $^{14}$N, as well as on the isospin-symmetry-breaking (ISB) correction 
to the superallowed $0^+ \rightarrow 0^+$ $\beta$ decay of $^{10}$C. We demonstrate, among other findings, that within the 
DFT-NCCI framework CI has a negligible effect on the ground-state isospin impurities, which are dominated by a single doorway state. 
In contrast, CI significantly modifies the impurities in excited states, including the isobaric analogue $I=0^+,\,T=1$ state in $^{10}$B. 
Hence, it also has a non-negligible impact on the ISB correction to the superallowed $\beta$ decay of $^{10}$C. Our calculations 
yield $\bar{\delta}_{\rm C}=0.45(4)\%$ when the Coulomb interaction is taken as the sole source of ISB, 
and $\bar{\delta}_{\rm ISB}=0.46(6)\%$  when short-range charge-symmetry-breaking (CSB) terms are included in addition. 
Hence, no statistically significant dependence of the ISB correction on the short-range CSB interaction is observed for this decay. 
Comparison with our previous results reveals a strong sensitivity to the nuclear symmetry energy, which governs the strength of 
the isospin-restoring force and whose value in finite nuclei remains difficult to constrain because of its intricate dependence on the 
momentum-dependent terms of the effective interaction.
}
\end{abstract}

\maketitle

\section{Introduction} 

Isospin-symmetry breaking (ISB) originates from the unequal masses of the up and down quarks, which generate the strong component of the symmetry 
breaking, and from their different electric charges, which give rise to the electromagnetic contribution that dominates in atomic nuclei. Within effective 
theories, where quarks and gluons are replaced by interacting nucleons or, at an even more phenomenological level, by the densities and currents they 
generate, the physical picture becomes considerably more complex. In such approaches, ISB emerges from several intertwined mechanisms, including the 
neutron--proton mass difference, one- and two-boson exchange processes (such as $2\pi$ exchange with intermediate $\Delta$ excitations, as well as 
$\pi\rho$ and $\rho\omega$ exchanges), pion mass splitting, and $\pi\gamma$ exchange contributions.

Such interactions, including those constructed on purely phenomenological grounds, can be grouped into three distinct categories introduced by Henley and Miller~\cite{(Hen79)}: class-II (isotensor or charge-independence-breaking, CIB), class-III (isovector or charge-symmetry-breaking, CSB), and class-IV forces . 
These classes of interactions manifest themselves in specific many-body observables, often in an almost exclusive manner. For example, mirror and triplet 
displacement energies (MDEs and TDEs) are primarily sensitive to class-III and class-II forces, respectively, whereas differences in neutron and proton 
analyzing powers in neutron--proton scattering provide evidence for class-IV interactions.

A prerequisite for all ISB-related phenomena is isospin mixing. The probability of admixture of subleading isospin components in a 
nuclear many-body wave function -- commonly referred to as  the isospin or Coulomb impurity ($\alpha_{\rm C}$) -- provides 
a natural measure of the degree of ISB in atomic nuclei.  The quantity $\alpha_{\rm C}$ is, however, not directly accessible experimentally in a 
model-independent manner and therefore belongs to the class of so-called {\it pseudo-observables\/}.

The first theoretical estimates of $\alpha_{\rm C}$ date back to the mid-1950s and are due to MacDonald. Assuming that the effect originates entirely 
from the isovector part of the Coulomb interaction and is dominated by coupling to the lowest $T=1$ state, he derived a perturbative expression for 
an upper bound on $\alpha_{\rm C}$ in the $T_0=0$ ground states of light nuclei~\cite{(MacD55),(MacD56)}:
\begin{equation}
\alpha_{\rm C}^{\rm (max)} = \frac{A(A-1)}{2}\frac{(e^2/R)^2}{8(E_0-E_1)^2}(0.76 + 0.017A),
\end{equation}
where $R$ denotes the nuclear radius, while $E_0-E_1$ is the excitation energy of the lowest $T_0+1$ state. Since this quantity is not determined 
within the model, it was taken -- somewhat arbitrarily -- from experimental data, typically in the range of 10-20\,MeV.
Under this {\it ad hoc\/} assumption, MacDonald's pioneering numerical estimates indicated that ISB impurities in light
 $N=Z$ nuclei (up to $^{20}$Ne) are  typically of the order of one percent i.e. indeed very small.  His approach was subsequently generalized to 
 the ground states of heavier nuclei, as well as to nuclei with $T_0\neq 0$, in Ref.~\cite{(LS62)}.

 In 1965, Sliv and Kharitonov~\cite{(SK65)} derived the first perturbative expression for $\alpha_{\rm C}$ within the independent-particle model. 
The central idea was to replace the two-body Coulomb interaction by an average Coulomb field which, inside a uniformly charged nucleus, can 
be approximated by a harmonic-oscillator potential. It is well known that such a potential induces particle-hole excitations between single-particle 
states of the same parity, characterized by $\Delta j=\Delta \ell=0$ and $\Delta n=1$. Consequently, the dominant $T_0+1$ configuration 
-- hereafter referred to as the {\it doorway state} -- is formed by excitations across two major oscillator shells and is therefore located at relatively high 
excitation energies of the order of
\begin{equation}
\Delta E=2\hbar\omega, \, \hbar\omega=\frac{41}{A^{1/3}}\, {\rm MeV}.
\end{equation}

The inclusion of the symmetry-energy term in the nuclear mean field, as proposed by A. Bohr {\it et al.}~\cite{(Boh67)},
\begin{equation}\label{Esym}
V_{\rm sym} = \frac{V_1}{A} {\bm t}\cdot {\bm T}, \quad V_1 \approx 100\,{\rm MeV},
\end{equation}
leads to a further reduction of $\alpha_{\rm C}$. Indeed, the symmetry energy acts as an isospin-restoring force, thereby suppressing 
isospin mixing or, equivalently, shifting the $T_0+1$ doorway state to higher excitation energies:
\begin{equation}\label{S-K-2}
\Delta E=2\hbar\omega \rightarrow 2\hbar\omega + \frac{V_{1}}{A} (T_0+1).
\end{equation}

Within the independent-particle model, the calculated isospin impurities do not exceed a few percent and decrease approximately as $\alpha_{\rm C}\propto 1/(T_0+1)$ with increasing distance from the $N\approx Z$ line. More realistic treatments of the Coulomb interaction, such as that of Ref.~\cite{(KW69)}, lead to similar estimates of isospin mixing.

Although small, these impurities are nevertheless larger than those predicted by the hydrodynamical model of Bohr {\it et al.}~\cite{(Boh67)}, which represents one of the earliest attempts to incorporate collective correlations. In that model, the symmetry force acts as the restoring force for monopole oscillations of the isovector density. As a consequence, the strength associated with the $T_0+1$ doorway state is shifted to substantially higher excitation energies,
\begin{equation}
\Delta E = \frac{169}{A^{1/3}}\, \rm{MeV},
\end{equation}
leading to an even stronger suppression of isospin mixing. Summarizing, the early models emphasized the role of Coulomb interaction 
as a primary source of ISB in atomic nuclei and the role of symmetry energy which, acting 
as a symmetry-restoring force, was reducing $\alpha_{\rm C}$. These models were disregarding the role of correlations as well as
the impact of configuration mixing with low -- or intermediate -- energy $T_0+1$ states. 

As this brief historical overview of isospin impurity illustrates, isospin symmetry and its breaking have attracted the attention 
of generations of physicists for several decades. In recent years, the field has experienced renewed interest, driven by increasingly 
precise experimental measurements and the development of new theoretical concepts and models aimed at describing various 
aspects of isospin-symmetry-breaking (ISB)-sensitive observables; see, for example, 
Refs.~\cite{(Ben07),(Tow10),(Kan17),(Roc18),(Har20),(Lle20),(Mar21),(Nai22),(Xay22),(Sen23),(Hoc25),(Zim25)} 
and the references therein.

The aim of this work is to investigate the impact of configuration mixing  on $\alpha_{\rm C}$  and on the isospin-symmetry-breaking 
(ISB) correction to the superallowed Fermi $0^+\rightarrow 0^+$ decay of $^{10}$C using the  framework based on multi-reference DFT 
theory involving configuration-interaction and rigorous treatment of both rotational and isospin symmetries which was developed by our group specifically 
with the aim of studying ISB phenomena~\cite{(Sch12),(Sch17),(Dob21)}. The framework was successfully applied to compute 
$\alpha_{\rm C}$~\cite{(Sat09),(Sat10)} and ISB corrections to the superallowed $0^+ \rightarrow 0^+$~\cite{(Sat11),(Sat12)} 
and $T=1/2$ mirror~\cite{(Kon22)} $\beta-$decays. Over the last  decade we  extended the formalism by including the No-Core 
Configuration-Interaction  (DFT-NCCI) module~\cite{(Sat16d)} and generalizing the Skyrme force by including zero-range class II (isotensor) 
and class III (isovector) forces  up to the next-to-leading order (NLO) in the gradient expansion~\cite{(Bac18),(Bac19)}. These allowed 
us to extend the range of applications to cover mirror- and triplet-energy displacements (MDEs and TDEs)~\cite{(Bac18),(Bac19)} as 
well as  mirror- and triplet-energy differences (MEDs and TEDs)  along rotational bands in mirror nuclei~\cite{(Bac21),(Zim25)} and, 
in turn, further increase a credibility of our modeling of  the ISB phenomena.

The DFT-NCCI configuration space is constructed incrementally by incorporating relevant (multi)particle–(multi)hole configurations step by step. 
This provides a unique opportunity to introduce the low- and intermediate-energy $T=T_0+1$ states into the model space in a controlled, sequential 
manner and to investigate their impact on the isospin mixing in the $T_0$ ground state (g.s.). In particular, it enables a direct assessment of early theoretical 
interpretations that attributed the effect to a single doorway state.

In our study we will focus on calculating  $\alpha_{\rm C}$ for three representative  $N\approx Z$ cases: ({\it i\/}) near-spherical nucleus $^{14}$N 
and ({\it ii\/}) well deformed nuclei $^{10}$B and  $^{10}$C. Moreover, we shall address a role of correlations, configuration mixing and 
strong-force charge-symmetry-breaking forces on the ISB correction to the superallowed   $^{10}$C$\rightarrow^{10}$B decay 
which is within reach of various many-body techniques including {\it ab initio\/} approaches.

The paper is organized as follows. In Sect.~\ref{sec:ncci} we concisely introduce the concept of DFT-NCCI 
model. In Sect.~\ref{sec:imp} we discuss the isospin impurity in the  g.s. and the excited  
isospin doublet of $^{14}$N and the lowest $T=1,I=0$ state of $^{10}$B. In Sect.~\ref{sec:10C} we present 
the results of calculations of $\delta_{\rm C}$ for the superallowed $0^+\rightarrow 0^+$ decay of $^{10}$C. 
The paper is briefly summarized in Sec.~\ref{sec:sum}

\section{The DFT-NCCI approach}\label{sec:ncci}

The self-consistent Hartree–Fock(-Bogoliubov) approach provides an efficient framework for describing complex nuclear-structure 
phenomena in terms of intuitive deformed single-particle configurations. Owing to its conceptual simplicity and predictive power, 
it forms the foundation of many beyond-mean-field theories that incorporate correlations between the mean-field configurations. 
A characteristic feature of single-reference nuclear density functional theory (SR-DFT) is the spontaneous breaking of fundamental 
symmetries. While this mechanism is essential for an accurate description of bulk nuclear properties, it limits the applicability of the 
method to quantities such as excitation spectra and electromagnetic or weak transition rates. This limitation can be overcome by 
restoring the broken symmetries through projection techniques, thereby extending SR-DFT to the multi-reference DFT (MR-DFT) framework. 
In MR-DFT, the nuclear wave function is constructed as a superposition of symmetry-transformed Slater determinants, 
$\hat R |\varphi\rangle$, where $\hat R$ denotes rotations in coordinate, isospin, or gauge space and the corresponding weights are 
dictated by the underlying symmetry group. Additional mixing of states projected from different particle-hole or quasiparticle configurations 
further enhances the descriptive power of the method, bringing it to a level comparable with that of the nuclear shell model (NSM).

Several implementations of DFT-based configuration-interaction approaches have been proposed; for a recent overview see Ref.~\cite{(She21)}. 
In this work we will use the DFT-NCCI based on an unpaired Skyrme energy density functional.  The method combines angular-momentum and
isospin projection techniques in a unified manner, specifically tailored for studies of isospin-symmetry breaking (ISB) in nuclei with $N\approx Z$. 
Without entering into the details, the model's computational scheme proceeds in the following three major steps: 
\begin{enumerate}
\item
The first step is the construction of the {\it configuration space}. To this end, we determine a set of $N_{\rm conf}$ relevant low-lying 
self-consistent Hartree–Fock configurations, including particle-hole and multiparticle-multihole excitations,
$\{ |\varphi_j\rangle \}_{j=1}^{N{\rm conf}}$. These reference states provide the basis for the subsequent 
symmetry-restoration procedure.
\item
In the second stage, 3D angular-momentum and 1D isospin projections are applied to each Hartree–Fock configuration 
$\ket{\varphi}\in \{ |\varphi_j\rangle \}_{j=1}^{N{\rm conf}}$. 
The resulting states are:
\begin{equation}\label{IMKT}
|\varphi; \, IMK; \, T T_z\rangle 
    = \frac{1}{\sqrt{N_{\varphi; IMK; TT_z}}}  \hat P^T_{T_z T_z} \hat P^I_{MK} |\varphi \rangle,
\end{equation}
where
\begin{eqnarray}
  \hat P^T_{T_z, T_z}  &\,& =  \nonumber \\  &\,& \frac{2T+1}{2}   \int_0^\pi
d^{T}_{T_z T_z}(\beta_T ) e^{-i\beta_T \hat{T}_y}
 \sin\beta_T\, d\beta_T,   \\
  \hat P^I_{M, K} &\,& =  \nonumber  \\  &\,& \frac{2I+1}{8\pi^2 }   \int
 D^{I\, *}_{M K}(\Omega ) 
 e^{-i\gamma \hat{J}_z} e^{-i\beta \hat{J}_y} e^{-i\alpha \hat{J}_z}
 \, d\Omega, 
\end{eqnarray}

are the standard 1D isospin and 3D angular-momentum projection operators respectively \cite{(RS80),(Var88)}.
Since $K$ and $T$ are not conserved  a subsequent $K$- and $T$-mixing procedure is carried out at this stage, 
leading to a set of good-angular-momentum states free from unphysical isospin mixing
forming the so called {\it model space\/}:
\begin{eqnarray}\label{m-space}
\ket{\varphi_j ; I M; T_z}^{(i)} &\,& =\\ \frac{1}{\sqrt{\mathcal{N}^{(i)}_{\varphi_j ; IM; T_z}}} &\,&
\sum_{\substack{K, T \geq |T_z|}} a_{K, T}^{(i)} \, \hat P^T_{T_z T_z}  \hat{P}^{I}_{MK}\ket{\varphi_j}.
\end{eqnarray} 
\item
In the final step, the Hamiltonian is diagonalized in the {\it model space\/}.   
As a result, a set of linearly independent DFT-NCCI eigenstates is obtained in the form:
\begin{equation}
\ket{\psi_{\textrm{NCCI}}^{k; IM; T_z}}
=\frac{1}{\sqrt{\mathcal{N}^{(k)}_{IM;T_z}}}
\sum_{i,j}c_{ij}^{(k)}\, \ket{\varphi_j;I M; T_z}^{(i)}\,, \label{eq:nccistate}
\end{equation}  
where $k$ enumerates eigenstates in ascending order according to their energies. 
The NCCI quantum states are free from spurious  isospin mixing. 
\end{enumerate}

The final diagonalization (as well as the $K-$ and $T-$mixing) must be performed 
with care due to overcompleteness of the {\it model space\/}. In the HFODD code 
(see~\cite{(Dob09d)} for further details) we  handle these problems 
by solving the Hill-Wheeler-Griffin (HWG) in the 
{\it collective space\/} spanned by linearly independent {\it natural states\/}:
\begin{equation}   \label{nat_st}   |IM; T_z\rangle^{(m)} =
  \frac{1}{\sqrt{n_m}} \sum_{ij} \eta_{ij}^{(m)}
  {\ket{\varphi_j ; I M; T_z}^{(i)}}.
  \end{equation}
These {\it natural states} are constructed from the eigenstates 
of the norm matrix:
\be\label{egn:eignorm}
\sum_{i'j'} N_{ij ; i'j'} \bar{\eta}^{(m)}_{i'j'} = n_m \bar{\eta}^{(m)}_{ij},
\ee
where, for reasons of numerical stability, only the eigenstates having eigenvalues 
$n= 1,2,...,m_{\text{max}}$ satisfying $n_m > \zeta$ — with $\zeta$ being a 
user-defined basis cut-off parameter — are retained. In the calculations presented below, 
we use typically $\zeta = 0.01$. Similar procedure is applied to  $K-$ and $T-$mixing.

The calculations presented here were performed using a developing version of the HFODD 
solver~\cite{(Dob09d),(Sch17),(Dob21)}, which has been equipped with a DFT-NCCI module.
In the calculations presented below, we use a spherical basis consisting either 10 or 
12 harmonic oscillator shells. Integration over the Euler angles 
$\Omega = (\alpha, \beta, \gamma)$  is performed using Gauss-Chebyshev quadrature 
(for $\alpha$ and $\gamma$) and Gauss-Legendre quadrature (for $\beta$ and $\beta_T$), with 
$n_\alpha = n_\beta = n_\gamma = 20$ knots. Integration over the Euler angle $\beta_T$ in isospace 
was done using $n_{\beta_T} =8$ knots.   

In the present calculations, while computing configurations, we superimpose parity, time-reversal (for the even-even 
nucleus $^{10}$C) and the signature symmetry: 
\begin{equation}\label{signature}
    \hat{R}_y = e^{-i \pi \hat{J}_y},
\end{equation}
defined as a rotation by an angle $\pi$ about $y-$axis. Consequently, the single-particle states are eigenstates 
of $\hat{R}_y$ and are labeled by multiplicative -- in the many-body context -- eigenvalues $r = \pm i$.

In the calculations presented hereafter we use, depending on the case, three different variants of the 
density-independent isoscalar Skyrme pseudo-potential  SV of Ref.~\cite{(Bei75)} [albeit always with tensor terms 
included in the SV-EDF for the sake of mathematical consistency]. The variants include: the SV$_{\rm{SO}}$ force
with increased by 20\% the spin-orbit strength, see~\cite{(Kon16)}; the SV$_{\rm{SO;D}}$ having the 
symmetry-energy strength adjusted to the isobaric binding-energy parabolas:
\begin{equation}\label{BE_parabolas} 
BE(T_z)=a +  b T_z + c T_z^2;
\end{equation} 
and the  SV$_{\rm{SO;D}}^{\rm NLO}$  
augmented  with class~III CSB interaction in the next-to-leading (NLO) order:
\begin{eqnarray}
\hat{V}^{\rm{III}}(i,j)  =  \bigg[&&
t_0^{\rm{III}}  \delta\left(\gras{r}_{ij} \right)
  +  \frac12 t_1^{\rm{III}}
\left( \delta\left( \gras{r}_{ij} \right) \bm{k}^2 + \bm{k}'^2 \delta\left(\gras{r}_{ij} \right) \right)   
\nonumber \\
& &+  t_2^{\rm{III}}
\bm{k}' \delta\left(\gras{r}_{ij} \right) \bm{k} \bigg]  \left( \hat{\tau}_3^{(i)}+\hat{\tau}_3^{(j)} \right) ,
\label{eq:classIII_NLO}
\end{eqnarray}
where $\gras{r}_{ij} = \gras{r}_i - \gras{r}_j$,
$\bm{k}  =  \frac{1}{2i}\left(\bm{\nabla}_i-\bm{\nabla}_j\right)$ and
$\bm{k}' = -\frac{1}{2i}\left(\bm{\nabla}_i-\bm{\nabla}_j\right)$ are the standard relative-momentum
operators acting to the right and left, respectively.
The three new  LECs are  equal: $t_0^{\rm{III}}= {}$\mbox{$5\pm2$\,MeV\,fm$^3$},
$t_1^{\rm{III}}= {}$\mbox{$-3\pm3$\,MeV\,fm$^5$}, and $t_2^{\rm{III}} = {}$\mbox{$-7.4\pm0.7$\,MeV\,fm$^5$}.
They have been adjusted globally to all available data on MDEs for $A\geq 6$ in Refs.~\cite{(Bac19),(Dob21)}, which makes   
our approach free from adjustable parameters.


\section{The DFT-NCCI analysis of the isospin impurity in $^{14}$N}\label{sec:imp}

\begin{figure}[ht!]
\centering
\includegraphics[scale=0.50, clip]{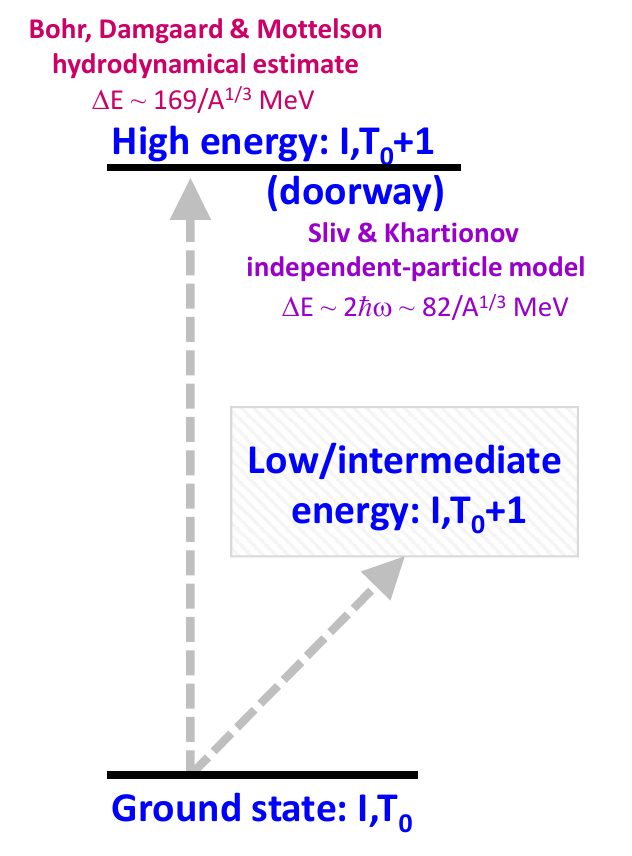}
\caption{(Color online) Schematic illustration of possible isospin mixing 
sources in the g.s. considered in this work: the high-energy doorway and low- 
or intermediate-energy $T=T_0+1$ states. 
} 
\label{fig:impurity}
\end{figure}

According to the early analytical models~\cite{(SK65),(Boh67)}, the isospin impurities in the g.s. wave functions  were attributed 
to the mixing with a single high-energy doorway state having isospin $T_0+1$. In complex nuclei, however, low- and intermediate-energy 
$T_0+1$ states may appear and contribute non-negligibly to the isospin mixing, as schematically illustrated in Fig.~\ref{fig:impurity}. In this section, 
we examine the validity of the single-state dominance scenario in the near-spherical nucleus $^{14}$N. The analysis will be extended in the next 
section to the well-deformed nucleus $^{10}$B, where, however, we shall focus on the impurities to the lowest $I=0, T=1$ state.

The $^{14}$N nucleus provides an excellent testing ground for investigating the relative roles of a doorway state and low-/intermediate-energy $T=T_0+1$ 
states in generating isospin mixing, owing to the simplicity of its configuration space. Because $^{14}$N is a nearly spherical nucleus, it is more natural 
and intuitive to describe its mean-field configurations in terms of spherical quantum numbers, which we adopt throughout the following discussion.

Although we employ a spherical notation, it is important to emphasize that, unlike in the Nuclear Shell Model (NSM), the mean-field configurations break 
both rotational and isospin symmetries. Consequently, they implicitly incorporate couplings to high-energy excitations that do not need to be included 
explicitly in the configuration space. This allows us to restrict the configuration space to elementary $1p-1h$ and $2p-2h$ excitations within the $p$ 
shell, supplemented by a few lowest $2p-2h$ excitations to the $sd$ shell to test stability of our predictions. 

To facilitate the discussion, it is convenient to divide the configurations into the following four groups:
\begin{itemize}
\item 
{\bf Group 1} consists of the two lowest configurations representing the g.s. and corresponding to the parallel (aligned) and antiparallel 
(antialigned) couplings of the unpaired neutron and proton, namely, $\nu p_{1/2}\otimes\pi p_{1/2}$ and $\nu p_{1/2}\otimes\widetilde{\pi p_{1/2}}$,
respectively. Here, $\widetilde{\pi p_{1/2}}$ denotes the signature-reversed partner of the $\pi p_{1/2}$ state. The complementary configuration,
$\widetilde{\nu p_{1/2}}\otimes\pi p_{1/2}$, obtained by reversing the neutron signature instead, is equivalent to 
$\nu p_{1/2}\otimes\widetilde{\pi p_{1/2}}$ and can therefore be omitted.
\item
{\bf Group 2} includes the three (i.e., all) $1p-1h$ proton excitations within the $p$ shell, ${\nu p_{1/2}}\otimes\pi p_{3/2}^{-1}$.
\item
{\bf Group 3} consists of the isobaric analogs, $\nu p_{3/2}^{-1}\otimes\pi p_{1/2}$, of the configurations belonging to Group 2.
\item
{\bf Group 4} contains all $2p-2h$ configurations within the $p$ shell, $\nu p_{3/2}^{-1}\otimes\pi p_{3/2}^{-1}$.
\end{itemize}

\begin{figure}[t!]
\centering
\includegraphics[scale=0.45]{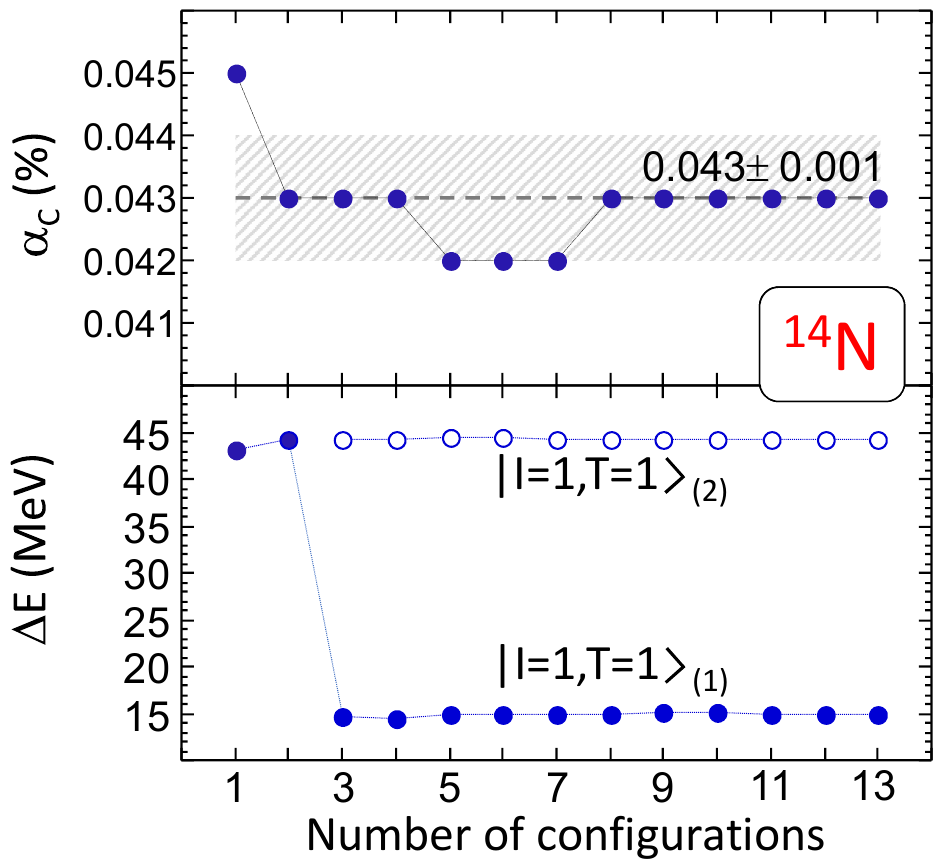}
\caption{(Color online) The upper panel shows the isospin mixing, $\alpha_{\rm C}$, in the g.s., $|I=1,T\approx 0\rangle$, of $^{14}$N 
as a function of the number of configurations included in the DFT-NCCI calculations. The lower panel shows the excitation energies of the doorway 
state and, beginning with the lowest configuration of Group 2, the lowest $|I=1,T\approx 1\rangle_1$ state built from 
$1p$-$1h$ excitations within the $p$ shell. The configurations are ordered by groups and, within each group, arranged in ascending order of their 
excitation energies. 
} 
\label{fig:alpha14N}
\end{figure}

The isospin mixing, $\alpha_{\rm C}$, in the g.s., $|I=1,T\approx 0\rangle$, of $^{14}$N is shown in the upper 
panel of Fig.~\ref{fig:alpha14N} as a function of the number of configurations included in the DFT-NCCI calculations. The configurations 
are grouped according to their type and, within each group, ordered by increasing excitation energy.

For the two lowest configurations, the entire isospin mixing originates from a single doorway state located at an excitation energy of approximately 
45\,MeV; see the lower panel of Fig.~\ref{fig:alpha14N}. This state emerges naturally within the model through the long-range Coulomb polarization 
effect. Its excitation energy agrees well with the independent-particle-model estimate given by Eq.~(\ref{S-K-2}), which predicts a value of about 
41\,MeV.

The inclusion of the lowest $1p-1h$ excitation from Group 2 (and/or Group 3) gives rise to a $|I=1,T\approx 1\rangle$ state located at an excitation 
energy of approximately 15\,MeV, that is, nearly three times lower than that of the doorway state; see the lower panel of Fig.~\ref{fig:alpha14N}. 
Remarkably, this low-energy $T=T_0+1$ state has virtually no impact on the value of $\alpha_{\rm C}$ calculated for the g.s..
The same conclusion applies to the mixing with the $2p-2h$ configurations belonging to Group 4. The present case thus provides a striking example 
demonstrating that the single-doorway-state dominance mechanism proposed in Refs.~\cite{(SK65),(Boh67)} works almost perfectly.

\begin{figure}[ht!]
\centering
\includegraphics[scale=0.45]{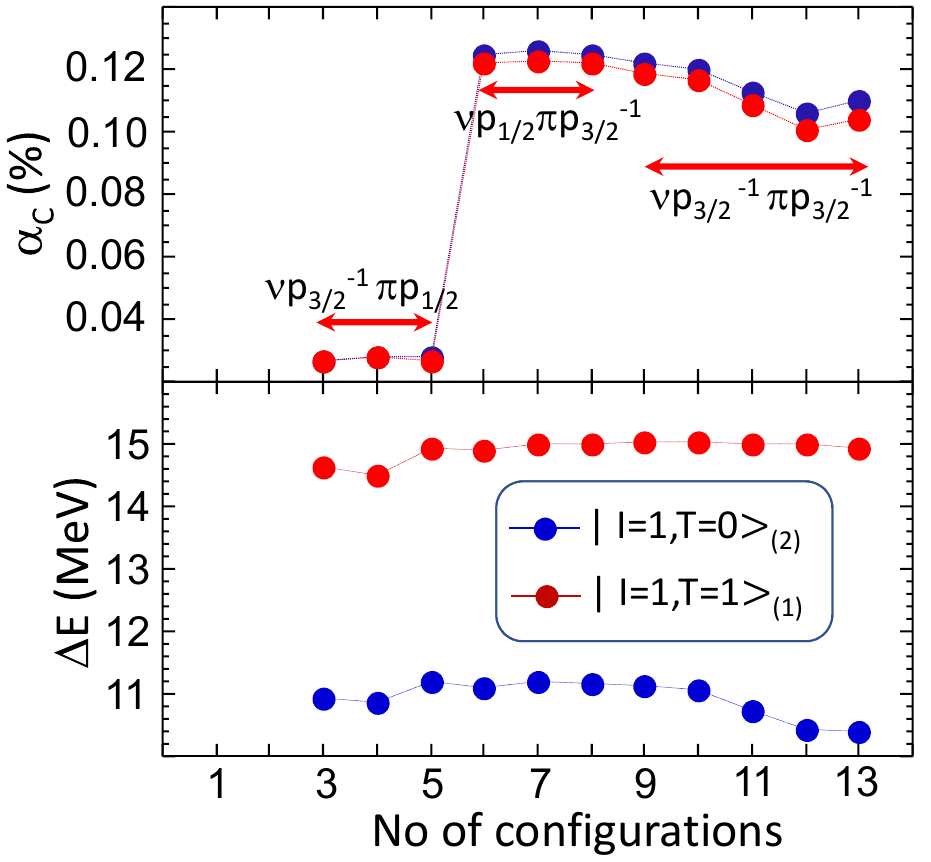}
\caption{(Color online) The upper panel shows the isospin mixing, $\alpha_{\rm C}$, in the $|I=1,T\approx 0\rangle$ and 
$|I=1,T\approx 1\rangle$ doublet in $^{14}$N built upon $1p-1h$ belonging to the Groups 2 and 3. 
The configurations are ordered by groups and, within each group, arranged in ascending order of their 
excitation energies. 
} 
\label{fig:doublet}
\end{figure}

The inclusion of $1p-1h$ excitations in the configuration space, in fact, leads to the appearance of a doublet of states, $|I=1^+,T\approx 0\rangle$ 
and $|I=1^+,T\approx 1\rangle$, having very similar structures but different isospins, as illustrated in Fig.~\ref{fig:doublet}. The $|I=1^+,T\approx 1\rangle$ 
state is shifted by approximately $\Delta E_{10}\approx 4$\,MeV  relative to the $|I=1^+,T\approx 0\rangle$ state due to the symmetry energy and 
can therefore serve as an independent measure of the strength of the symmetry-energy potential.

From the perspective of isospin mixing, these two states closely track each other as the size of the configuration space increases, indicating 
that they form an almost perfectly isolated doublet. The calculated isospin mixing amounts to $0.11\pm 0.01$\% and is fairly, although not entirely, 
insensitive to the configuration mixing within the $p$ shell. It is important to note, however, that a proper description of the isospin mixing requires 
the simultaneous inclusion of both the ${\nu p_{1/2}}\otimes\pi p_{3/2}^{-1}$ configuration and its isobaric analogue, 
$\nu p_{3/2}^{-1}\otimes\pi p_{1/2}$. In other words, isospin symmetry can be properly assessed within our model only if it is not explicitly 
violated by construction of the configuration space.

The analyzed isospin doublet should not be confused with the well-studied negative-parity doublet $|I=1^-,T\approx 0\rangle$ and 
$|I=1^-,T\approx 1\rangle$, see~\cite{(Ren72),(Kim22)} and the references therein, which arises from $1p$--$1h$ excitations 
across the $N=Z=8$ shell gap and therefore lies beyond the configuration space considered here.

\section{Superallowed Fermi $\beta$-decay of $^{10}$C}\label{sec:10C}

In this section, we present the results of our calculations of the isospin-symmetry-breaking (ISB) correction to the 
superallowed $0^+\rightarrow0^+$ Fermi $\beta$ decay of $^{10}$C. This transition is of particular importance 
because it is the lightest precisely measured superallowed emitter and plays a unique role in stringent tests of the 
Conserved Vector Current (CVC) hypothesis and the unitarity of the Cabibbo-Kobayashi-Maskawa (CKM) matrix. 
Owing to its low nuclear charge, the transition is especially sensitive to possible scalar weak currents beyond the 
Standard Model. Consequently, a reliable determination of the ISB correction is essential for interpreting high-precision 
experimental data and for maximizing the sensitivity of this decay to physics beyond the Standard Model.

In addition, the $^{10}$C decay can be addressed using a broad range of nuclear many-body methods, including state-of-the-art 
{\it ab initio\/} approaches. It therefore provides a unique benchmark for comparing and cross-validating different many-body 
techniques, offering valuable insight into their ability to describe the subtle interplay of Coulomb and strong-interaction effects 
responsible for isospin-symmetry breaking.

Unlike our previous studies, the present work systematically investigates, for the first time, the combined impact of configuration 
mixing and local charge-symmetry-breaking (CSB) interactions on the calculated ISB correction. This analysis allows us to assess 
the reliability of the theoretical predictions with respect to both the size of the configuration space and the treatment of the 
nuclear interaction.

Because both the isospin mixing and the resulting ISB corrections arise from a delicate interplay between the Coulomb interaction 
and the ISB components of the strong nuclear force, on the one hand, and the isospin-conserving nuclear interaction, which 
generates the isospin-restoring symmetry potential, on the other, an accurate description of the nuclear symmetry energy is 
a prerequisite for reliable calculations. We therefore begin by readjusting the symmetry-energy strength of the SV Skyrme 
interaction to reproduce the empirical systematics of finite nuclei before proceeding to the calculation of the ISB corrections.

\subsection{Adjustment of the symmetry energy - the SV$_{\rm SO;D}$ parametrization}\label{ssec:SVD}

Let us start by briefly recalling that for the Skyrme EDF, in the limit of uniform nuclear matter, the nuclear symmetry energy (NSE) is quadratic 
in the reduced isospin $I=(N-Z)/A$:
\begin{equation}\label{FGM}
E_{\rm sym}/A = a_{\rm sym} I^2
\end{equation}
The strength of the NSE equals,
\begin{equation}\label{inf}
   a_{\rm sym} = \frac{1}{8}\frac{m}{m^*} \varepsilon_{FG}
   + \left[    C_1^\rho \rho_0 + \left( \frac{3\pi^2}{2} \right)^{2/3} C_1^\tau \rho_0^{5/3}
   \right],
\end{equation}
where the first term in Eq.~(\ref{inf}) is the kinematic contribution, which can be associated with the average level spacing in the Fermi-gas 
model, $\varepsilon_{\rm FG}$, scaled by the isoscalar effective mass $m^*$, whereas the second term is interaction-driven and depends 
on the leading isovector contributions to the Skyrme EDF, namely $C_1^\rho \rho_1^2$ and $C_1^\tau \rho_1\tau_1$, where 
$\rho_1$ and $\tau_1$ denote the local isovector particle and kinetic-energy densities, respectively.

The term proportional to $C_1^\rho$ describes the isovector response of the central potential and depends on the overall strength of the 
central interaction, $t_0$, and the spin-exchange parameter, $x_0$, according to
\begin{equation}
C_1^\rho = -\frac{1}{4}t_0\left(\frac{1}{2}+x_0\right),
\end{equation}
for density-independent parametrizations. The contribution proportional to $C_1^\tau$ introduces a nontrivial dependence on 
momentum-dependent effects through the isovector effective mass. For details, we refer the reader to 
Refs.~\cite{(Ben03),(Sat03a),(Sat06),(Naz14)}.

\begin{figure}[t!]
\centering
\includegraphics[scale=0.45]{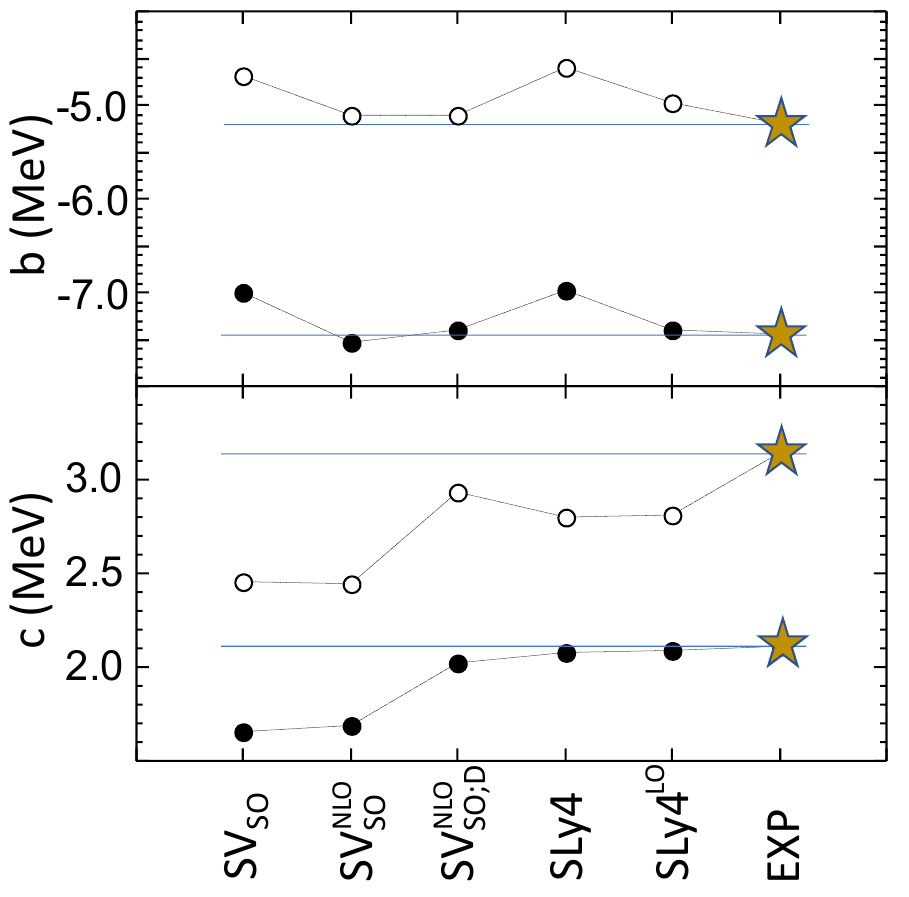}
\caption{(Color online) Upper (lower) panel shows the values of the parameters $b$ ($c$) of the isobaric mass parabola for the 
two selected isobaric chains, $A=23$ (open circles) and $A=40$ (filled circles). From left to right, the points correspond to 
calculations performed using the SV$_{\rm SO}$, SV$_{\rm SO}^{\rm NLO}$, SV$_{\rm SO;D}$, SV$_{\rm SO;D}^{\rm NLO}$, 
SLy4, and SLy4$^{\rm LO}$ parametrizations. The stars and horizontal lines represent the experimental value. 
We do not show neither experimental nor theoretical uncertainties of the parabolic fits for the sake of clarity of the figure. 
} 
\label{fig:b_and_c}
\end{figure}

The overall strength of the symmetry energy for the SV force in uniform nuclear matter is $a_{\rm sym}=32.8$\,MeV, in very good agreement with 
the empirical estimate $a_{\rm sym}^{\rm (exp)}=32\pm2$\,MeV. Despite this agreement, the SV force clearly underestimates the symmetry energy 
in finite nuclei, as can be inferred from direct calculations of the isobaric mass (binding energy BE) parabola (\ref{BE_parabolas}).
This deficiency is illustrated in Fig.~\ref{fig:b_and_c}, which shows that, for the SV$_{\rm SO}$ variant of the interaction, 
neither the $b$ nor the $c$ coefficient is reproduced satisfactorily. For clarity, the figure presents results for two representative 
isobaric chains, $A=23$ and $A=40$, selected from the four chains $A=23$, 34, 40, and 48 considered in this work.

In principle, the interaction should be refitted. However, the available data on light- and medium-mass $A\leq 100$ magic nuclei 
provide only weak constraints on the nuclear symmetry energy, since they are dominated by nuclei with $N\approx Z$; see 
Ref.~\cite{(Sat12)} for the results of such a refit. Instead of performing a complete refit of the Skyrme interaction, we therefore adopt 
a simplified approach. Specifically, we modify the symmetry-energy strength by varying the parameter $x_0$ incrementally, 
taking $x_0=-0.15$, $-0.10$, $-0.05$, and 0, and subsequently recalculate the coefficients $b$ and $c$ of the isobaric mass parabola 
to reproduce the experimental data. Although this procedure is admittedly somewhat ad hoc, it is expected to have only a limited 
impact on other aspects of nuclear structure because, apart from the symmetry energy, the parameter $x_0$ affects only the isoscalar 
time-odd term $C_0^s{\bm s}^2$ in the energy density functional, where ${\bm s}$ denotes the spin density.

As already noted, the calculations show that the SV$_{\rm SO}$ force fails to reproduce both the $b$ and $c$ coefficients. 
The SV$_{\rm SO}^{\rm NLO}$ interaction reproduces the $b$ coefficient without any additional readjustment of the CSB 
low-energy constants (LECs), but it still fails to account for the $c$ coefficient. A satisfactory description of the latter is 
obtained after setting $x_0=0$, leading to the interactions denoted SV$_{\rm SO;D}$ and SV$_{\rm SO;D}^{\rm NLO}$, 
as shown in Fig.~\ref{fig:b_and_c}. It is worth noting that the SLy4 interaction augmented with local CSB terms reproduces the 
isobaric mass parabola satisfactorily without any further adjustment of its LECs.

In the following we will therefore use the SV$_{\rm SO;D}$ and SV$_{\rm SO;D}^{\rm NLO}$ interactions what will allow 
to assess the impact of short-range CSB terms on the $\delta_\text{ISB}$ value. 

\subsection{Configuration space in $^{10}$C and $^{10}$B}\label{ssec:10CCONF}

Let us recall that, for even-even nuclei, we impose time-reversal symmetry, which restricts the available configurations in 
$^{10}$C to seniority-zero neutron-neutron and proton-proton pairing configurations. Moreover, in the present calculations 
we consider only axially symmetric configurations. Under these assumptions, the number of configurations involving active 
orbitals originating from the spherical $p$ shell is limited to nine. These configurations are shown schematically in 
Table~\ref{tab:10C_CONF}, where they are ordered according to increasing excitation energy.

The choice of the configuration space in $^{10}$C determines, through the requirement of isospin symmetry, 
the corresponding configuration space in $^{10}$B. In other words, to avoid introducing an explicit violation of isospin 
symmetry through the construction of the configuration space, the $^{10}$B basis must contain all configurations that 
are isobaric analogues of those included for $^{10}$C. A straightforward counting shows that this leads to a total of 17 
configurations, as listed in Table~\ref{tab:10B_CONF}, or 18 configurations when the two possible orientations of the 
ground-state configuration are taken into account. The ground-state configuration corresponds to a well-deformed prolate 
shape with a quadrupole deformation of $\beta_2\approx 0.38$. Its symmetry axis can be aligned either parallel or perpendicular 
to the signature axis $Oy$, which is the symmetry axis imposed in the calculations. In odd-odd nuclei, these two orientations 
are not equivalent because of the presence of time-odd mean fields and must therefore both be included in the 
configuration space. For all excited configurations, however, we retain only a single orientation, with the symmetry axis 
aligned parallel to the signature axis $Oy$. Their role should be reduced due to excitation energy or, alternatively, the weight 
in the wave function and also because they are less deformed.

\begin{table}[h!]
\centering
\caption{(Color online)  
Configurations used in the DFT-NCCI calculations for $^{10}$C. Blue (red) dots$^*$ denote pairwise occupied neutron (proton) 
single-particle (s.p.) states labeled by their asymptotic Nilsson quantum numbers. The corresponding spherical subshells from 
which the Nilsson orbitals originate are also indicated for clarity. Owing to time-reversal symmetry, only seniority-zero 
neutron-neutron and proton-proton pairing-like excitations are considered.  
}
\label{tab:10C_CONF}
\begin{tabular}{|c|cc|cc|cc|cc|cc|cc|cc|cc|cc|}\hline 
\rule{0pt}{0.35cm} configuration: & \multicolumn{2}{c|}{$C^{\rm C}_1$} & \multicolumn{2}{c|}{$C^{\rm C}_2$} & \multicolumn{2}{c|}{$C^{\rm C}_3$} & \multicolumn{2}{c|}{$C^{\rm C}_4$} & \multicolumn{2}{c|}{$C^{\rm C}_5$} & \multicolumn{2}{c|}{$C^{\rm C}_6$} & \multicolumn{2}{c|}{$C^{\rm C}_7$} & \multicolumn{2}{c|}{$C^{\rm C}_8$} & \multicolumn{2}{c|}{$C^{\rm C}_9$} \\\hline   
$d_{5/2}$; [220\,1/2] & 
& &
& &
& &
& &
& &
& &
& &
& &
&  \\  
$p_{1/2}$; [101\,1/2] & 
                                               &                          &
                                               &                          &
                                               & \tikzcircle{0.7ex} &
                                               & \tikzcircle{0.7ex} &
\tikzcircle[blue, fill=blue]{0.7ex} &                          &
                                               & \tikzcircle{0.7ex} &
                                               & \tikzcircle{0.7ex} &
\tikzcircle[blue, fill=blue]{0.7ex} & \tikzcircle{0.7ex} &
\tikzcircle[blue, fill=blue]{0.7ex} & \tikzcircle{0.7ex}    \\  
$p_{3/2}$; [101\,3/2] & 
                                               & \tikzcircle{0.7ex} &
\tikzcircle[blue, fill=blue]{0.7ex} & \tikzcircle{0.7ex} &
                                               &                          &
\tikzcircle[blue, fill=blue]{0.7ex} & \tikzcircle{0.7ex} &
                                               & \tikzcircle{0.7ex} &
                                               & \tikzcircle{0.7ex} &
\tikzcircle[blue, fill=blue]{0.7ex} &                          &
                                               &                          &
                                               & \tikzcircle{0.7ex}    \\  
$p_{3/2}$; [110\,1/2]  & 
\tikzcircle[blue, fill=blue]{0.7ex} & \tikzcircle{0.7ex} &
                                               & \tikzcircle{0.7ex} &
\tikzcircle[blue, fill=blue]{0.7ex} & \tikzcircle{0.7ex} &
                                               &                          &
                                               & \tikzcircle{0.7ex} &
\tikzcircle[blue, fill=blue]{0.7ex} &                          &
                                               & \tikzcircle{0.7ex} &
                                               & \tikzcircle{0.7ex} &
                                               &                              \\\hline  
\end{tabular}
\end{table}
\vspace{-1.3em}
\begin{minipage}{0.90\textwidth}
\footnotesize
\textsuperscript{*} $\tikzcircle[blue, fill=blue]{0.7ex} := {\color{blue} \bm\uparrow}{\color{blue} \bm\downarrow},\ \tikzcircle[red, fill=red]{0.7ex} := {\color{red} \bm\uparrow}{\color{red} \bm\downarrow}$
\end{minipage}

\begin{table}[h!]
\centering
\caption{(Color online)  
Configurations used in the DFT-NCCI calculations for $^{10}$B. Full dots denote pairwise occupied 
s.p. states while the up (down) arrows denote singly occupied s.p. states with positive (negative) $K$ 
quantum numbers, respectively. Blue/red symbols refer to neutrons/protons, respectively. 
For convenience, the states are labeled using both the spherical and the asymptotic Nilsson quantum numbers. 
Only the isospin anoluge configurations with respect to the seniority zero pp/nn pairing-like excitations in $^{10}$C 
are included.  
}
\label{tab:10B_CONF}
\begin{tabular}{|c|cc|cc|cc|cc|cc|cc|cc|cc|}\hline 
\rule{0pt}{0.35cm} configuration: & \multicolumn{2}{c|}{$C^{\rm B}_{1/2}$} & \multicolumn{2}{c|}{$C^{\rm B}_3$} & \multicolumn{2}{c|}{$C^{\rm B}_4$} & \multicolumn{2}{c|}{$C^{\rm B}_5$} & \multicolumn{2}{c|}{$C^{\rm B}_6$} & \multicolumn{2}{c|}{$C^{\rm B}_7$} & \multicolumn{2}{c|}{$C^{\rm B}_8$} & \multicolumn{2}{c|}{$C^{\rm B}_9$} \\ \hline   
$d_{5/2}$; [220\,1/2] & 
& &
& &
& &
& &
& &
& &
& &
& \\  
$p_{1/2}$; [101\,1/2] & 
                                               &                                                 &
                                               &                                                 &
{\color{blue} $\bm\uparrow$}    & {\color{red} $\bm\downarrow$}  &
{\color{blue} $\bm\uparrow$}    & {\color{red} $\bm\downarrow$}  &
                                               & \tikzcircle{0.7ex}                        &
{\color{blue} $\bm\uparrow$}    & {\color{red} $\bm\downarrow$}  &
\tikzcircle[blue, fill=blue]{0.7ex} &                                                 &
{\color{blue} $\bm\uparrow$}    & {\color{red} $\bm\downarrow$}  \\  
$p_{3/2}$; [101\,3/2] & 
{\color{blue} $\bm\uparrow$}    & {\color{red} $\bm\downarrow$}  &
\tikzcircle[blue, fill=blue]{0.7ex} & \tikzcircle{0.7ex}                        &
                                               &                                                 &
\tikzcircle[blue, fill=blue]{0.7ex} & \tikzcircle{0.7ex}                        &
\tikzcircle[blue, fill=blue]{0.7ex} &                                                 &
{\color{blue} $\bm\uparrow$}    & {\color{red} $\bm\downarrow$}  &
                                               &  \tikzcircle{0.7ex}                       &
                                               &  \tikzcircle{0.7ex}                       \\  
$p_{3/2}$; [110\,1/2]  & 
\tikzcircle[blue, fill=blue]{0.7ex} & \tikzcircle{0.7ex}                        &
{\color{blue} $\bm\uparrow$}    & {\color{red} $\bm\downarrow$}  &
\tikzcircle[blue, fill=blue]{0.7ex} & \tikzcircle{0.7ex}                        &
                                               &                                                 &
{\color{blue} $\bm\uparrow$}    & {\color{red} $\bm\downarrow$}  &
{\color{blue} $\bm\downarrow$}& {\color{red} $\bm\uparrow$}      &
{\color{blue} $\bm\downarrow$}& {\color{red} $\bm\uparrow$}      &
\tikzcircle[blue, fill=blue]{0.7ex} &                                                 \\\hline\hline  
\rule{0pt}{0.35cm} configuration: & \multicolumn{2}{c|}{$C^{\rm B}_{10}$} & \multicolumn{2}{c|}{$C^{\rm B}_{11}$} & \multicolumn{2}{c|}{$C^{\rm B}_{12}$} & \multicolumn{2}{c|}{$C^{\rm B}_{13}$} & \multicolumn{2}{c|}{$C^{\rm B}_{14}$} & \multicolumn{2}{c|}{$C^{\rm B}_{15}$} & \multicolumn{2}{c|}{$C^{\rm B}_{16}$} & \multicolumn{2}{c|}{$C^{\rm B}_{17}$} \\\hline   
$d_{5/2}$; [220\,1/2] & 
& &
& &
& &
& &
& &
& &
& &
& \\  
$p_{1/2}$; [101\,1/2] & 
{\color{blue} $\bm\uparrow$}    & {\color{red} $\bm\downarrow$}  &
{\color{blue} $\bm\uparrow$}    & {\color{red} $\bm\downarrow$}  &
                                               & \tikzcircle{0.7ex}                        &
\tikzcircle[blue, fill=blue]{0.7ex} &                                                 &
{\color{blue} $\bm\downarrow$}& {\color{red} $\bm\uparrow$}      &
{\color{blue} $\bm\downarrow$}& {\color{red} $\bm\uparrow$}      &
\tikzcircle[blue, fill=blue]{0.7ex} &  \tikzcircle{0.7ex}                       &
\tikzcircle[blue, fill=blue]{0.7ex} &  \tikzcircle{0.7ex}                       \\  
$p_{3/2}$; [101\,3/2] & 
\tikzcircle[blue, fill=blue]{0.7ex} &                                                 &
{\color{blue} $\bm\uparrow$}    & {\color{red} $\bm\downarrow$}  &
{\color{blue} $\bm\downarrow$}& {\color{red} $\bm\uparrow$}      &
{\color{blue} $\bm\uparrow$}    & {\color{red} $\bm\downarrow$}  &
{\color{blue} $\bm\uparrow$}    & {\color{red} $\bm\downarrow$}  &
{\color{blue} $\bm\uparrow$}    & {\color{red} $\bm\downarrow$}  &
                                               &                                                 &
{\color{blue} $\bm\uparrow$}    & {\color{red} $\bm\downarrow$}  \\  
$p_{3/2}$; [110\,1/2]  & 
                                               & \tikzcircle{0.7ex}                        &
{\color{blue} $\bm\uparrow$}    & {\color{red} $\bm\downarrow$}  &
\tikzcircle[blue, fill=blue]{0.7ex} &                                                 &
                                               & \tikzcircle{0.7ex}                        &
{\color{blue} $\bm\downarrow$}& {\color{red} $\bm\uparrow$}      &                                               
{\color{blue} $\bm\uparrow$}    & {\color{red} $\bm\downarrow$}  &                                              
{\color{blue} $\bm\downarrow$}& {\color{red} $\bm\uparrow$}      &
                                               &                                                  \\\hline\hline  
\end{tabular}
\end{table}

\subsection{The results for isospin-symmetry breaking correction}\label{ssec:10CCONF}

The configurations can be conveniently labeled by specifying the occupation numbers of the active Nilsson orbitals:
$[n_1, n_2, n_3] \equiv [[101\, 1/2]^{n_1}, [101\, 3/2]^{n_2}, [110\, 1/2]^{n_3}]$. 
A one-to-one correspondence between the configurations in $^{10}$C and their analogs in $^{10}$B exists only for configurations in which 
an entire isospin quartet is occupied. This includes: 
$C_1^{\rm C} \leftrightarrow  C_{1/2}^{\rm B}$,  $C_2^{\rm C} \leftrightarrow  C_{3}^{\rm B}$, 
$C_3^{\rm C} \leftrightarrow  C_{4}^{\rm B}$, $C_4^{\rm C} \leftrightarrow  C_{5}^{\rm B}$,
$C_8^{\rm C} \leftrightarrow  C_{16}^{\rm B}$, and $C_9^{\rm C} \leftrightarrow  C_{17}^{\rm B}$. 
For the $[2,2,2]$ occupation, however, isospin symmetry requires the simultaneous inclusion of all three configurations, 
$\{ C_5^{\rm C},\, C_6^{\rm C},\, C_7^{\rm C} \}$,  in  $^{10}$C together with all nine corresponding configurations 
$C_i^{\rm B}$ where $i=6,\ldots 15$ in $^{10}$B.  Otherwise, the calculated ISB corrections may be contaminated by an artificial 
breaking of isospin symmetry arising from an asymmetric choice of the configuration spaces.

\begin{figure}[ht!]
\centering
\includegraphics[scale=0.45]{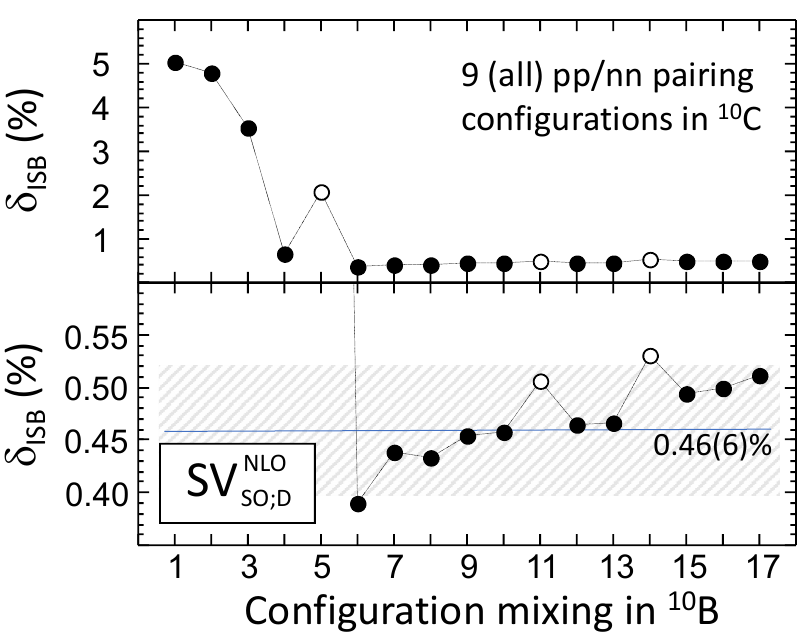}
\caption{(Color online) Stability of $\delta_{\rm ISB}$ in function of a number of configurations in $^{10}$B for
a fully correlated $^{10}$C. Open circles  indicate configurations causing instabilities in the excited states. These 
three configurations are removed from the calculations i.e. the points represented by black dots are free from their 
contributions.  Vertical line in the lower panel indicates the mean value 
$\bar{\delta}_{\rm ISB}= 0.46(6)$.
} 
\label{ISB_vs_10B}
\end{figure}

\begin{figure}[ht!]
\centering
\includegraphics[scale=0.45]{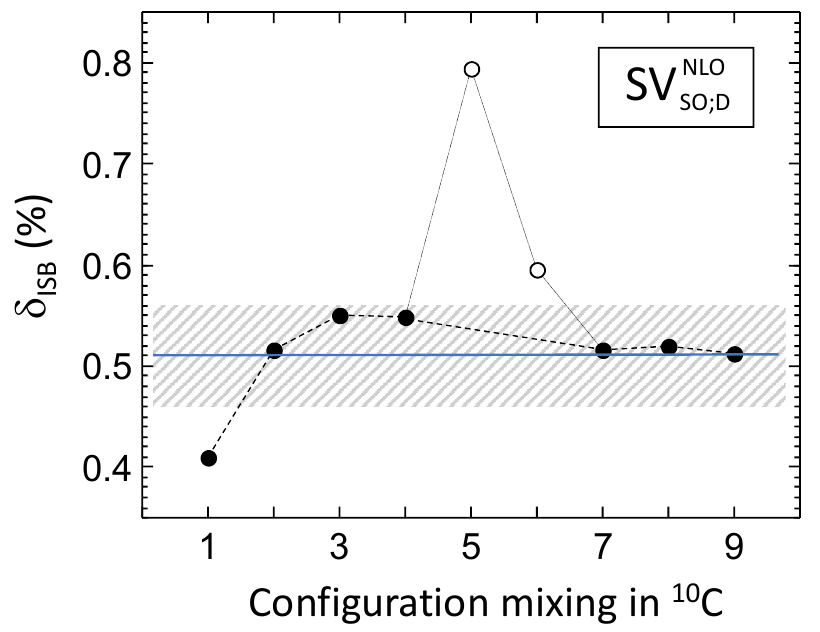}
\caption{(Color online) Stability of $\delta_{\rm ISB}$ in function of a number of configurations in $^{10}$C for
a fully correlated $^{10}$B. The calculated correction is almost perfectly stable $\delta_{\rm ISB} = 0.51(5)\%$. 
The open circles  illustrate the unphysical isospin breaking effect due to asymmetric choice of bases caused by 
adding the configurations $C_5^{\rm C}$ and $C_6^{\rm C}$.   
} 
\label{ISB_vs_10C}
\end{figure}

The effect is clearly illustrated in Figs.~\ref{ISB_vs_10B} and~\ref{ISB_vs_10C}, which demonstrate the stability of the 
calculated $\delta_{\rm ISB}$ with respect to the size of the configuration space. The symbols $\delta_{\rm C}$ ($\alpha_{\rm C}$) 
and $\delta_{\rm ISB}$ ($\alpha_{\rm ISB}$) are used interchangeably. The latter notation emphasizes that the calculated 
correction (or impurity) includes both Coulombic and non-Coulombic ISB contributions.
Figure~\ref{ISB_vs_10B} shows  $\delta_{\rm ISB}$ as a function of the number of configurations included in the $^{10}$B 
configuration space. In these calculations, we use the complete configuration space in $^{10}$C to describe the $I=0$, $T=1$ ground state, 
while the corresponding isobaric analogue in $^{10}$B is calculated by gradually enlarging its configuration space. 
Including only a few configurations in $^{10}$B results in a severe violation of isospin symmetry, yielding $\delta_{\rm ISB}$ 
values of the order of a few percent. As the configuration space is expanded, the calculated ISB correction rapidly stabilizes 
reaching finally a value of approximately $0.51\%$, as shown in the lower panel of Fig.~\ref{ISB_vs_10B}.

\begin{figure}[ht!]
\centering
\includegraphics[scale=0.45]{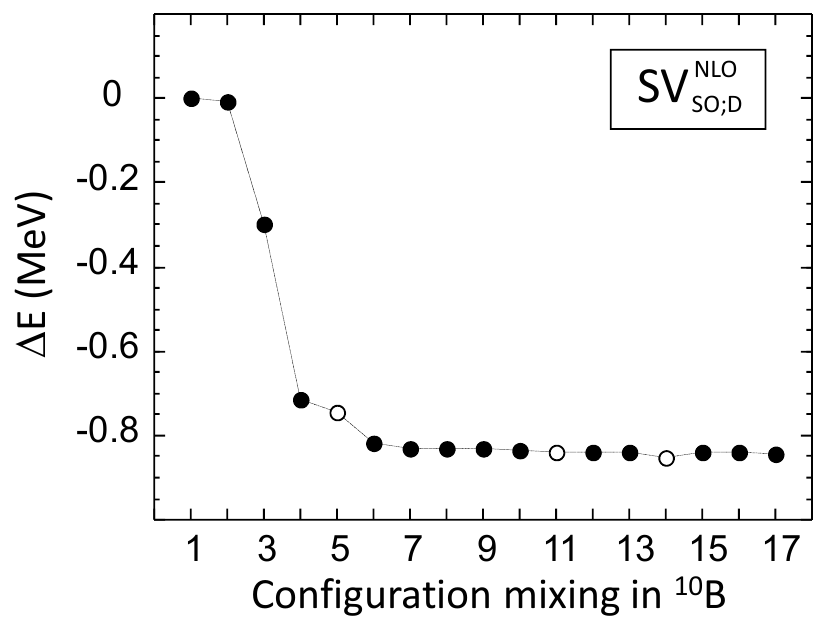}
\caption{(Color online) Energy gain of the lowest $|I=0^+, T=1\rangle$ state 
due to configuration mixing in $^{10}$B. Open circles indicate configurations which cause 
instabilities in the $\delta_{\rm ISB}$ calculations.  See text for further details.
} 
\label{e-gain}
\end{figure}

The three configurations marked by open circles exhibit instabilities in the excited states. The most pronounced instability, 
manifested by an anomalously low excitation energy of the projected $T=0$ state, is observed for configuration $C_5^{\rm B}$.
Although this configuration has only a minor effect on the energy of the $I=0, T=1$ state (see Fig.~\ref{e-gain}), it has a 
substantial impact on $\delta_{\rm ISB}$, which is extremely sensitive to even small admixtures of unfavored isospin components 
in the wave function. For this reason, we excluded this configuration from the $^{10}$B configuration space. The other two 
configurations exhibiting signs of instability, $C_{11}^{\rm B}$ and $C_{14}^{\rm B}$, were also excluded, although their effects 
appear to be much smaller and remain within the estimated theoretical uncertainty. The origin of these instabilities is not yet understood 
and requires further investigation

The ISB correction reaches a value of approximately 0.51(5)\% (with {\it ad hoc\/} 10\% uncertainty) when all reliable configurations 
are included. This value, however, may be artificially overestimated due to the unphysical ISB effect introduced by removing the three unstable 
configurations from the configuration space. In this sense, it may appear more appropriate to recommend as the final result the mean value, 
$\bar{\delta}_{\rm ISB}= 0.46(6)$, obtained by averaging over all black points shown in the lower panel 
of Fig.~\ref{ISB_vs_10B}. In this case, the quoted uncertainty has been increased relative to the standard deviation, 
$\sigma = 0.04$, to account for additional sources of uncertainty, such as the finite size of the harmonic-oscillator basis (10 shells are 
used in the present calculations) and the asymmetry of the configuration space resulting from the exclusion of the three uncertain 
configurations.

The effect of the unphysical breaking of isospin symmetry is also evident in Fig.~\ref{ISB_vs_10C}, which shows the stability of 
$\delta_{\rm ISB}$  as a function of the number of configurations included in the $^{10}$C configuration space. In these calculations, 
we use the complete $^{10}$B  configuration space (excluding the configurations that cause instabilities, as discussed above in 
connection with Fig.~\ref{ISB_vs_10B}) to describe the $I=0, T=1$ state, while the corresponding isobaric analogue in $^{10}$C
is obtained by gradually enlarging its configuration space. The unphysical symmetry breaking becomes clearly visible after the 
successive inclusion of the two members, $C_5^{\rm C}$ and $C_6^{\rm C}$ of the  $[2,\,2,\,2]$  triplet (open circles), which 
causes a sudden artificial increase in  $\delta_{\rm ISB}$. The inclusion of the third member of the triplet, $C_7^{\rm C}$, restores 
the artificially broken isospin symmetry and brings the calculated $\delta_{\rm ISB}$ back close to its average value.

\begin{figure}[ht!]
\centering
\includegraphics[scale=0.45]{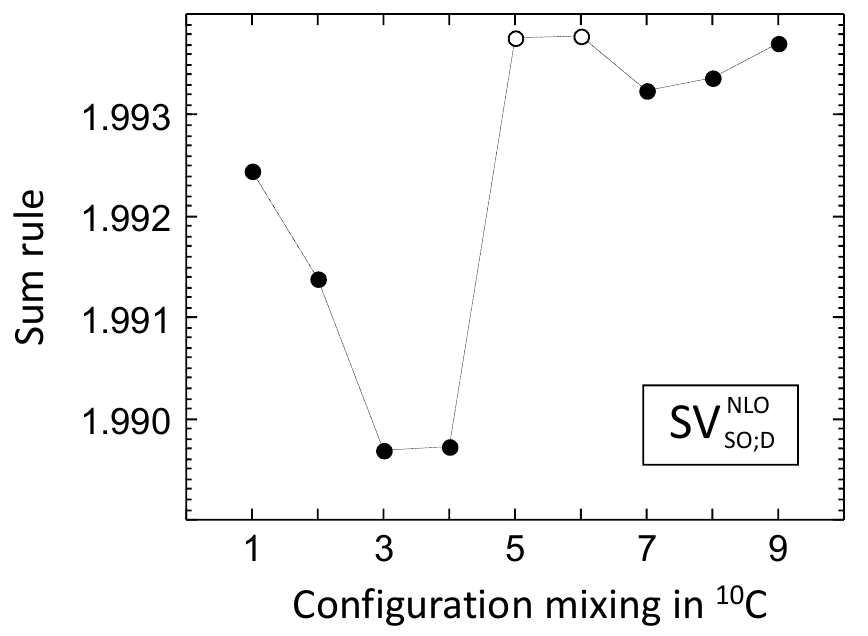}
\caption{(Color online) Fermi sum rule in function of the size of configuration space in $^{10}$C. 
The configurations are ordered according to the increasing excitation energy. 
All final $I=0$ states in $^{10}$B are included in the calculations at each point. 
} 
\label{sum_rule}
\end{figure}

Let us also point out that the Fermi sum rule defined as the sum of the reduced transition probabilities from the ground
state of $^{10}$C,  $\ket{\Psi_i}$, to all possible final states $\ket{\Psi_f}$ in $^{10}$B:
\begin{equation}
S_{\rm F} = \sum_f \left| \bra{\Psi_f} \hat{T}_+ \ket{\Psi_i} \right|^2,
\end{equation}
is fulfilled with very high accuracy as shown in Fig.~\ref{sum_rule}.  Indeed, the deviation from $S_{\rm F}=2$ corresponding to the ideal case 
is below 0.5\%.

\begin{figure}[ht!]
\centering
\includegraphics[scale=0.45]{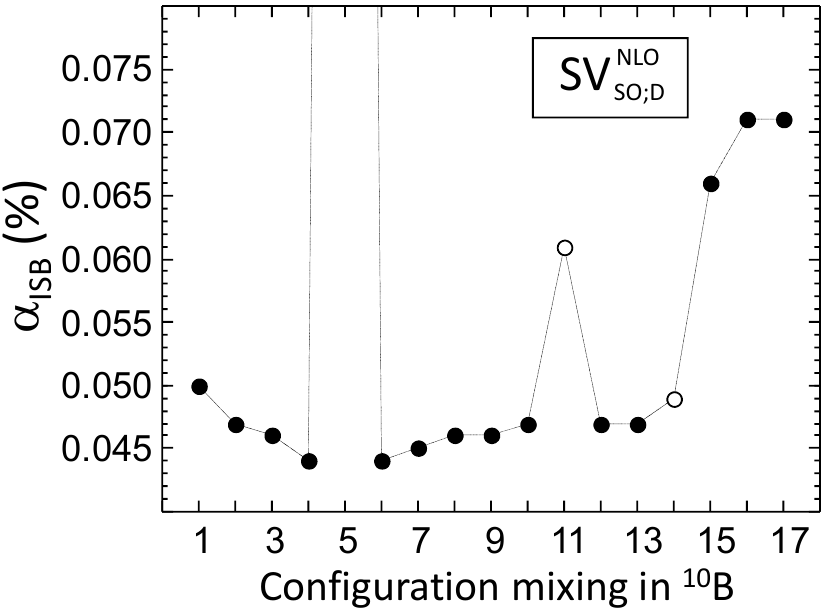}
\caption{(Color online) Isospin impurity $\alpha_{\rm ISB}$ in the lowest $|I=0^+, T=1\rangle$ state 
in $^{10}$B in function of the size of configuration space. The configurations are ordered according 
to the increasing excitation energy. Open circles indicate excluded configurations that
cause instabilities in the $\delta_{\rm ISB}$ calculations.  
} 
\label{alpha_C}
\end{figure}

It is important to emphasize that the final value of $\delta_{\rm ISB}=0.51(5)\%$, and,  to the lesser extent, 
the recommended value, $\bar{\delta}_{\rm ISB}=0.46(6)\%$, are noticeably larger than the result 
obtained using only the ground-state configurations, $C_1^{\rm C}$ and $C_{1/2}^{\rm B}$, which yields 
$\delta_{\rm ISB}=0.41(5)\%$. This observation suggests that configuration mixing plays a non-negligible role in determining 
the isospin impurity of the $|I=0^+, T=1\rangle$ state in $^{10}$B. This is indeed the case, as illustrated in Fig.~\ref{alpha_C}. 
The figure shows that the calculated isospin impurity of the $|I=0^+, T=1\rangle$ state in $^{10}$B increases with the number 
of configurations included in the DFT-NCCI model space, reaching its largest values when the highest-lying configurations are 
incorporated. This trend is nicely correlated with the increase of the calculated $\delta_{\rm ISB}$, although, as discussed above, part 
of this increase may originate from the asymmetry of the configuration spaces used for the parent and daughter nuclei.

We have performed a similar analysis of this decay using the SV$_{\rm SO;D}$ interaction and refrain from repeating the details here. 
To extract the final value in this case, we applied the same averaging procedure as that described above in connection with 
Fig.~\ref{ISB_vs_10B}, namely, by averaging $\delta_{\rm C}$ over the interval of relatively stable predictions and assigning a theoretical 
uncertainty of $10\%$ to the resulting average. This procedure yields $\bar{\delta}_{\rm C}=0.45(4)\%$.

This result indicates that the inclusion of CSB terms in the nuclear EDF does not lead to a statistically significant change in the 
predicted value of $\delta_{\rm C}$, in contrast to the analysis of the superallowed $T=1/2$ decays reported in Ref.~\cite{(Kon22)}.

Compared with our earlier study of $0^+ \rightarrow 0^+$ superallowed decays (Ref.~\cite{(Sat12)}), the present value of $\delta_{\rm C}$ 
is considerably smaller than the result obtained with the SV functional, $\delta_{\rm C}=0.65(14)\%$, but is consistent with the value 
$\delta_{\rm C}=0.462(65)\%$ calculated using the SHZ2 functional, which was adjusted to describe light nuclei with $A\leq100$. 
The SHZ2 functional is characterized by a significantly stiffer symmetry energy, $a_{\rm sym} \approx 42$\,MeV, than most commonly 
used Skyrme functionals, including SV, for which $a_{\rm sym} \approx 32$\,MeV. Its symmetry energy is, however, comparable 
to that of the SV$_{\rm SO;D}$ functional employed in the present work. It is worth to add that the confidence-level test (CL) proposed 
by Towner and Hardy in Ref.~\cite{(Tow10)} gives for the decay the so called (semi-)empirical value of the ISB correction   
$\delta_{\rm C}^{{\rm (exp)}}=0.37(14)\%$, see Ref.~\cite{(Sat12)}, what seems to contradict the 
SV result $\delta_{\rm C}=0.65(14)\%$ in spite of sizable uncertainties.

Let us note that the anomalously large value of $\delta_{\rm C}=0.65(14)\%$ obtained with the SV functional originates 
from a difference in the intrinsic shapes of the mean-field ground-state solutions for $^{10}$C and its mirror partner $^{10}$B. 
While the $^{10}$C solution is triaxial, the corresponding $^{10}$B solution is axially symmetric. Such a shape mismatch 
is exceptional and was found only in the $A=10$ system. It was interpreted as a manifestation of configuration mixing already 
at the mean-field level and provided one of the main motivations for undertaking the present configuration-interaction study.

\section{Summary and conclusions}\label{sec:sum}

Low-energy nuclear processes including the superallowed $0^+ \rightarrow 0^+$~\cite{(Har20)} 
and $T=1/2$ mirror~\cite{(Nav09a),(Gon19)} beta decays allow for precision tests of fundamental symmetries.
The tests heavily rely on precision calculations of ISB corrections being a domain of many-body 
nuclear models. Hence, precision of nuclear tests of the Standard Model is heavily intertwined
with the credibility of nuclear modeling of ISB phenomena.   

In this work we have presented the seminal DFT-NCCI calculations of isospin impurities in $^{14}$N and the 
ISB correction to the superallowed $0^+ \rightarrow 0^+$ \, $^{10}{\rm C}\rightarrow ^{10}{\rm B}$ \, $\beta$-decay.
 
Our calculations confirm, among the others, a single doorway state dominance for isospin impurity in the $I=1^+,\, T=0$ 
g.s. of nearly spherical odd-odd nucleus  $^{14}$N.  Similar single doorway state dominance holds also for 
the isospin impurity in the well $I=0^+,\, T=1$ deformed ground state of  $^{10}$C,  where our calculations give
$\delta_{\rm ISB}=0.009(1)\%$.  These two cases are in qualitative agreement with the early ideas based on independent-particle 
models~\cite{(SK65),(Boh67)}. In contrast, the isospin impurity in the isobaric analogue  $I=0, T_0=1$ state of $^{10}$B shows 
clear dependence on the mixing as shown  in Fig.~\ref{alpha_C}. In this case, however, the mixing comes from both the $T=0$ and 
$T=2$ states.  The CI also influences noticeably the isospin impurities in the $I=1^+$ isospin doublet in $^{14}$N.

Concerning the $\delta_{\rm C}$, our DFT-NCCI results are systematically larger than calculated values obtained using alternative 
models yielding $\bar{\delta}_{\rm C}=0.45(4)\%$ when the Coulomb interaction is taken as the sole source of ISB, 
and $\bar{\delta}_{\rm ISB}=0.46(6)\%$ when short-range CSB terms are included in addition. 
In particular, the perturbative expression of  Damgaard, see Refs.~\cite{(Dam69a),(Tow77),(Tow10)}, gives:
\begin{equation}
\delta_{\rm C} = 0.002645
\frac{Z^2}{A^{2/3}} (n + 1)(n + \ell + 3/2) (\%), 
\end{equation}
where $n$ and $\ell$ denote the number of radial nodes and angular momentum of the valence s.p. spherical wave function,
respectively, gives $\delta_{\rm C} = 0.046\%$. The shell-model plus Woods-Saxon (SM-WS) method of Ref.~\cite{(Har20)} yields
 0.175(18)\%. This value is, however, inconsistent with the  CL test yielding $\delta_{\rm C}\approx 0.33(15) \%$, 
 as can be inferred from Fig.~1 of Ref.~\cite{(Tow10)}. Recently,  Xayavong {\it et al.\/} \cite{(Xay25)} published new estimate 
 using a refined version of  the SM-WS approach $\delta_{\rm C} = 0.329(62)\%$. The relativistic Hartree or Hartree-Fock with RPA 
 calculations of \cite{(Lia09)} yield, for example, 0.088\% (for RHF-RPA with PKO3 Lagrangian) and 0.150\% (for RH-RPA with DD-ME2 
 Lagrangian). The isovector-monopole-resonance (IVMR) model of Auerbach \cite{(Aue09)} gives only 0.008\%. Recently, Piarulli {\it et al.\/}
\cite{(Pia26)} have communicated the seminal {\it ab initio\/} study of the decay utilizing quantum Monte Carlo methods. 
Their calculations, depending on the interaction, yield $\delta_{\rm C}$ values in the range 0.15\%-0.25\% with sizable relative 
uncertainties of order of 34\%-65\%. Hence, no statistically significant dependence on the choice of interaction can be inferred.

\begin{figure}[ht!]
\centering
\includegraphics[scale=0.45]{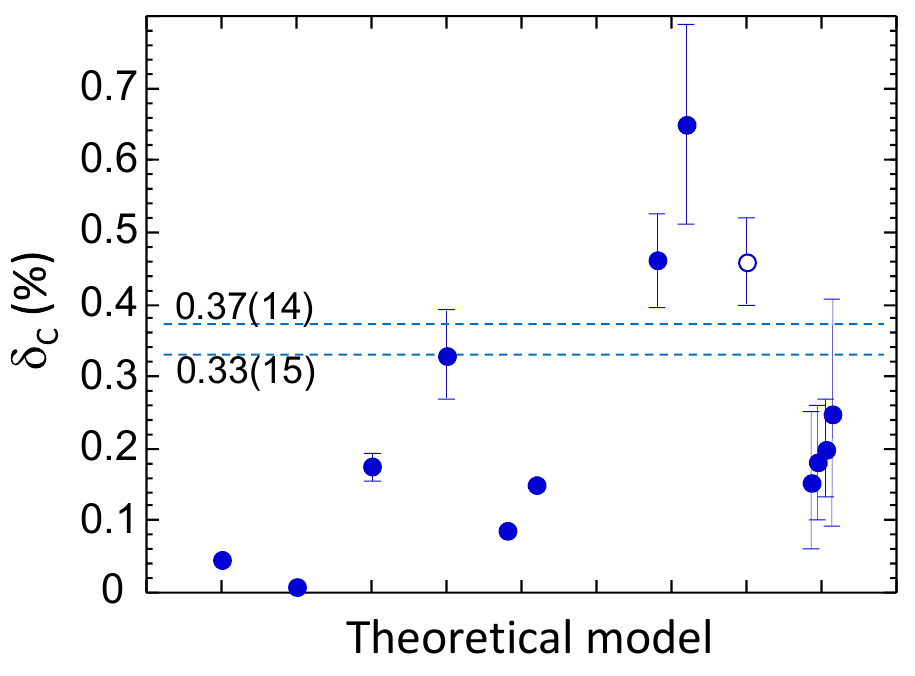}
\caption{(Color online) Survey of calculated isospin-symmetry-breaking corrections, $\delta_{\rm C}$ (or $\delta_{\rm ISB}$), 
for the superallowed $\beta$ decay of $^{10}$C obtained within various theoretical approaches (filled circles). From left to right, the results 
correspond to the Damgaard estimate~\cite{(Dam69a)}, the IVMR calculation of Ref.~\cite{(Aue09)}, the SM-WS result of Ref.~\cite{(Har20)}, 
the SM-WS result of Ref.~\cite{(Xay25)}, the RHF-RPA calculation based on the PKO3 Lagrangian and the RH-RPA calculation based on the 
DD-ME2 Lagrangian from Ref.~\cite{(Lia09)}, the DFT results of Ref.~\cite{(Sat12)} obtained with the SHZ2 and SV functionals, 
respectively, the DFT-NCCI result of the present work (open circle) , and the {\it ab initio\/} result of Ref.~\cite{(Pia26)}. The vertical lines 
indicate the pseudo-empirical central values inferred from the CL tests reported in Refs.~\cite{(Tow10)} and~\cite{(Sat12)}; see 
the text for additional details.  
} 
\label{delta_all}
\end{figure}

The available results for $\delta_{\rm C}$ are summarized in Fig.~\ref{delta_all}. The wide spread of the calculated values 
indicates that the superallowed $0^+ \rightarrow 0^+$ $^{10}{\rm C}\rightarrow ^{10}{\rm B}$ decay is surprisingly difficult 
to describe accurately. Consequently, even after excluding results obtained using schematic or certain less reliable models, 
the current precision of the calculated ISB corrections appears insufficient to support high-precision searches for physics 
beyond the Standard Model using this decay, in particular exotic scalar decays. Our calculations suggest that these difficulties 
may be related to the pronounced deformation of the ground states of both $^{10}{\rm C}$ and $^{10}{\rm B}$, which 
exhibit well-deformed prolate shapes with a quadrupole deformation parameter of approximately $\beta_2 \approx 0.38$.

Our calculations indicate that the ISB correction for the superallowed decay of $^{10}$C exhibits no statistically significant sensitivity 
to the short-range CSB interaction, in contrast to the ground-state Fermi decays in $T=1/2$ nuclei reported in Ref.~\cite{(Kon22)}. 
A comparison with our earlier results of Ref.~\cite{(Sat11)} reveals a pronounced sensitivity of the calculated correction to the 
nuclear symmetry energy, which determines the strength of the isospin-restoring force. Constraining the symmetry energy in finite 
nuclei remains challenging because of its intricate dependence on the momentum-dependent terms of the effective interaction, see 
Ref.~\cite{(Sat03a)}. 

 \vspace{0.5cm}
 
\begin{acknowledgments}

This work was supported by the Polish National Science Centre (NCN) (under Contract No 2018/31/B/ST2/02220) [JS and WS], and by the UK Science and Technology Funding Council (grant number ST/Y000323/1) [JS].

\end{acknowledgments}

\bibliographystyle{apsrev4-2}

\bibliography{ISB-NCCI,jacwit34}

\end{document}